\documentclass[letterpaper]{article}

\usepackage{ucs}
\usepackage[utf8x]{inputenc}
\usepackage[T1]{fontenc}

\usepackage{aaai}
\usepackage{times}
\usepackage{helvet}
\usepackage{courier}
\usepackage{dcolumn}
\usepackage{graphicx}
\title{Responses to remixing on a social media sharing website}

\author{Benjamin Mako Hill \\ Massachusetts Institute of Technology \\ 77 Massachusetts Avenue \\ Cambridge, MA 02139
\And Andr\'{e}s Monroy-Hern\'{a}ndez \\ Media Laboratory \\ Massachusetts Institute of Technology \\ 75 Amherst St \\ Cambridge, MA 02142
\And Kristina R. Olson \\ Department of Psychology\\ Yale University\\ 2 Hillhouse Ave.\\ New Haven, CT 06520}

\begin{document}
\maketitle

\begin{abstract}
In this paper we describe the ways participants of the Scratch online community, primarily young people, engage in remixing of
each others' shared animations, games, and interactive projects.
In particular, we try to answer the
following questions: How do users respond to remixing in a social media
environment where remixing is explicitly permitted?  What qualities of
originators and their projects correspond to a higher likelihood of
plagiarism accusations?  Is there a connection between plagiarism
complaints and similarities between a remix and the work it is based
on? Our findings indicate that users have a very wide range of reactions
to remixing and that as many users react positively as accuse remixers
of plagiarism.  We test several hypotheses that might explain the high
number of plagiarism accusations related to original project complexity,
cumulative remixing, originators' integration into remixing practice, and remixee-remixer
project similarity, and find support for the first and last explanations.
\end{abstract}

\section{Introduction}

Digital and networked information technologies have facilitated {\em
remixing}, the creation of new content, such as songs and video, using
existing content produced by others.  Manovich
\shortcite{manovich_remixability_2005} argues that remixing has taken
such a widespread and central role in contemporary
Internet culture that has it become, ``practically a built-in feature of
digital networked media universe.'' Benkler
\shortcite{benkler_wealth_2006} describes remixing as a fundamental
aspect of ``peer production'' and argues that if, ``we are to make this
culture our own, render it legible, and make it into a new platform for
our needs and conversations today, we must find a way to cut, paste, and
remix present culture.''

In the broadest sense, remixing is as old as creativity. Newton's famous
quote, ``if I have seen further it is only by standing on the shoulders
of giants,'' speaks to the importance of ```remixing'' others' ideas.
Benkler argues that new information technology and the Internet has, through
radically decreasing the costs of copying, distribution, modification, and collaboration,
increased the importance of remixing as a phenomenon enormously. Although these
authors, and others, have spoken to the importance and widespread nature
of remixing, there is very limited empirical research describing
the practice.



\begin{figure}
\begin{center}
\includegraphics[width=3in]{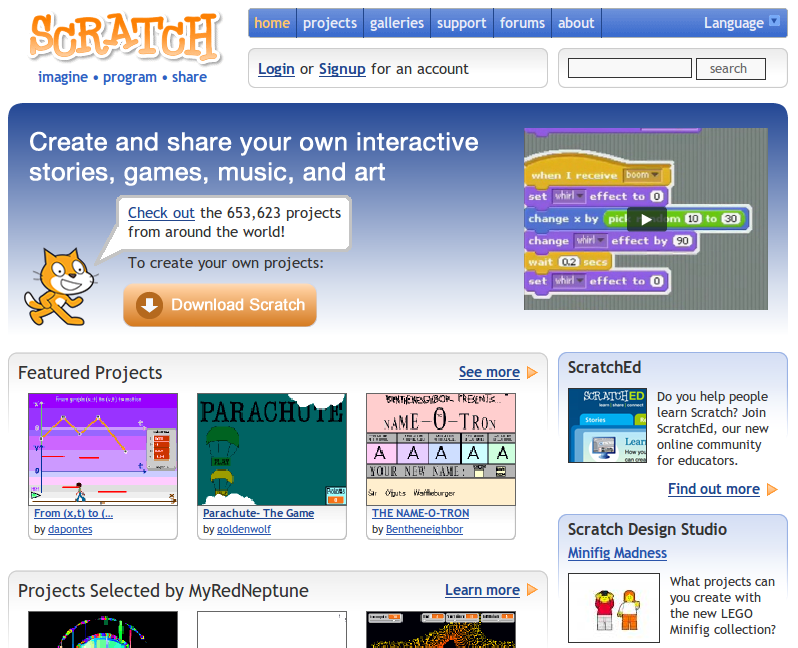}
\caption{Screenshot of the front page of the Scratch online community.}
\label{fig:frontpage}
\end{center}
\end{figure}

Much discussion of remixing has focused on the ways young people engage
in remixing. Jenkins \shortcite{jenkins_convergence_2006} and Ito
\shortcite{ito_personal_2006} have written extensively about children's
engagement in media remixing.  For example, Jenkins
\shortcite{jenkins_convergence_2006} has written about the way that
youth have played a pivotal role in online fan fiction communities to
rework characters and events in mainstream media franchises like Harry
Potter. These remixers have produced full length novels distributed
online, edited each others' work, and tutored each other. In their book,
{\em Born Digital}, Palfrey and Gasser \shortcite{palfrey_born_2008}
argue that remixing by ``digital natives'' is, ``already
having an effect on cultural understanding around the world.'' Jenkins
\shortcite{jenkins_confrontingchallenges_2006} argues that remixing or
``appropriation'' is so widespread and of such social significance, that
it should be among a set of core new media literacy skills taught in
schools. Lessig \shortcite{lessig_2007} has implored audiences to
prevent ``copyright extremism'' from ``strangling'' both children and
adults' amateur creativity and has proposed legal outlets to foster
``remix culture.''


Despite this interest in remixing on the Internet, researchers, like
Cheliotis and Yew \shortcite{cheliotis_analysis_2009} have pointed to
a dearth of empirical research on the topic. Cheliotis and Yew's
own study of the online remixing community {\em ccMixter} focused on
issues of remixer motivation and the way that contests can frame and
limit contributions. Other work has looked at {\em Jumpcut}, a video
sharing community, where Diakopoulos et al.
\shortcite{diakopoulos_evolution_2007} present interviews with remixers
that speak to the existence of unwritten community norms around
remixing. In two papers, Seneviratne et al.
\shortcite{seneviratne_detecting_2009,seneviratne_policy-aware_2009}
discuss the way that photos from the website {\em Flickr} are reused and
highlight issues of license compliance during reuse, showing that that
the majority of people do not give the proper attribution when using
{\em Flickr} photos on blog posts. Shaw and Schmitz
\shortcite{shaw_community_2006} describe a system used by the San
Francisco International Film Festival to facilitate remixing and
describe, both quantitatively and qualitatively, what users produced.

Although these studies represent important steps toward building an
empirical understanding of remixing, work to date has focused only on
why users remix, what remixing consists of, and how it may be done in
compliance with licenses. With the exception of {\em ccMixter}, each of
the remixing communities studied involves users creating remixes
primarily from raw materials created outside the community in question
although research has suggested that this distinction may be important.
For example, Diakopoulos et al. described users uploading film and
television snippets in clear violation of copyright while remaining wary
of remixing the work of other {\em Jumpcut} users. In this sense, much
previous work largely eschews discussion of the type of collaborative,
iterative, or cumulative collaborative work described by Benkler
\shortcite{benkler_wealth_2006} as peer production or by Murray and
O'Mahony \shortcite{murray_exploringfoundations_2007} as ``cumulative
innovation.'' Finally, all of the studies described have focused
primarily in remixing culture in adult communities despite the central
role that young people play.

In peer production or cumulative innovation, users are primarily engaged
in remixing each others' work and ideas. The result, as Diakopoulos et
al. begin to show in their qualitative descriptions, is the emergence of a complex
set of social dynamics that frame and motivate users' decisions to
collaborate and share. While Seneviratne et al. discuss the way that
systems to enforce licensing might be built, implicit in their
assumption is that strongly enforced license terms are important for
promoting sharing.  Even this basic assumption remains largely untested
empirically.

This paper represents an attempt to measure and explain reactions of
users to remixing in {\em Scratch}, a large social media sharing
community participated in primarily by young people. Although all
projects shared on Scratch are released under a license that permits
remixing, and although explicit Scratch community norms make it clear
that remixing is always permitted and encouraged, we show that there is
a wide range of reactions to remixing in Scratch.

This article presents three studies. The first study investigates the
range of reactions to remixing by the creators of projects that are
remixed and shows that although many users react positively to remixing,
an equal number accuse remixers of plagiarism. The second study
describes a logistic regression model that explores support for several
potential explanations for the large number charges of plagiarism based
on differences in antecedant projects and their authors.
Our final study makes an initial attempt at
establishing the role that the divergence between remixes and their
antecedents may play in originators' responses to being remixed. 

\section{Scratch}

The Scratch online community is a public website (See Figure
\ref{fig:frontpage}) where people from around the world share animated
stories, interactive art, and video
games.\footnote{\texttt{http://scratch.mit.edu}} Participants use the
{\em Scratch} programming environment \cite{resnick_scratch:_2009}, a
desktop application, to create projects by putting together images,
music and sounds with programming command blocks.

Two and a half years after the website was launched publicly in 2007,
close to 100,000 people had shared more than 800,000 Scratch projects.
Projects have ranged from interactive greeting cards to fractal
simulations to animations of lip-synching lizards to video games
featuring Obama and McCain.  Every month, about half a million people
visit the website to browse and interact with projects and other
users.\footnote{
\texttt{http://quantcast.com/scratch.mit.edu}.} Any visitor can browse
the website but must register for an account in order to upload or download
projects, comment, show support (``love''), tag other
users' projects, or to flag projects as inappropriate. There have been
nearly 360,000 accounts created by people with self-reported ages
ranging primarily from 8 to 17 years old, with 12 being the mode. Thirty-five percent of users are female.

A primary goal of the Scratch online community is to
encourage collaboration and remixing. All projects
shared on the website are licensed under the Creative Commons (CC)
Attribution Share-Alike License\footnote{
    \texttt{http://creativecommons.org/licenses/by-sa/2.0/}} and a link to a
child-friendly version of the license is displayed alongside every
project.  Users are able and encouraged to download any project, look
at or modify its graphics and code, and create new versions --- referred
to as remixes on the Scratch website --- which can then be shared themselves.

\section{Study 1: How do people respond to remixing?}

In Study 1, we assess Scratch users' reactions to having their projects
remixed. To do so, we created an algorithm to
identify all original-remix pairs in the first
13 months of the Scratch community. These pairs were made up of all
projects based on another project and its corresponding antecedent.  We
subsequently coded all comments left by the author of the original
project (i.e., the {\em originator}) on each remixed project.

Our algorithm identified projects based on metadata embedded in projects. As a result, we
would not count projects that were conceptually copied by a user who had
seen another's work but who did not actually copy code, graphics or
sounds.  Additionally, as there were over 100,000 projects, we could not
feasibly watch and interact with each project and determine whether the
original projects were actually ``original'' or whether ideas were taken
from a source outside Scratch (e.g., a user may have created a Pacman
clone).


\subsection{Procedure}

We applied our algorithm to all 136,929 projects created and posted on the
Scratch website between the community launch date in March 2007 and
April 2008 to identify all derivative projects;
a total of 11,861 projects were deemed to be based on other projects.
The comments left by originators on the first
3,555 projects were coded by two independent coders who were asked to
code all projects into the following categories: {\em no comment}
(projects in which the originator did not leave any comments on the
remixer's project), {\em positive} (projects in which the originator
left positive comments, e.g., ``Love what you did with my code! Great
idea!''), {\em hinting plagiarism} (projects in which the originator
implied that the remixer had copied but did not state this explicitly,
e.g., ``I mostly pretty much made this whole entire game''), {\em
plagiarism} (projects in which the the originator directly accused the
remixer of copying, e.g., ``Hello mr plagiarist”, ``Copy-cat!''), {\em
negative} (projects with negative comments that were not necessarily
related to copying, e.g., ``Alright you crap eating thumb sucking
baby''), and {\em none of the above} (projects with comments that were
not positive, negative, or relevant to plagiarism, e.g., ``is this
jarred'' or ``b for peanut butter jelly time!'').

The first and last categories were mutually exclusive from all others.
The other categories were potentially overlapping (an originator could
say, ``you copied me but I like your addition of the flowers'' and
therefore count as {\em positive} comments and {\em plagiarism}), with
the exception of {\em hinting plagiarism} and {\em plagiarism} which were mutually exclusive of each other. The coders were found to be reliable (absolute
agreement by category: {\em no comment}=100\%; of those that did include
a comment, coders agreed on the presence or absence of the following
categories at the following rates (absolute agreement): {\em
positive}=87\%, {\em hinting plagiarism}=81\%, {\em plagiarism}=89\%,
{\em negative}=93\%, {\em none of the above}=89\%). Therefore the
remaining comments (n=8,306) were split between the two coders, with
each coding approximately half of the remaining projects. For the few
projects that the coders disagreed on in the initial third of the
projects, they met and came to an agreement on the coding.

\subsection{Results}

Out of the 11,861 projects that were categorized by the algorithm as remixes,
we were able to determine that 3,742 (31.5\%) of the original creators
clicked on and saw the remixed versions of their projects. Of
those originators who saw the projects, 2,156 (58\%) did not leave a
comment.  Of those who saw the projects and commented, 261 (7\%) accused
the remixer of plagiarism, 566 (15\%) hinted at plagiarism concerns, 797
(21\%) left positive comments, 260 (7\%) left negative comments, and 237
(6\%) left comments that did not fit into these categories or were
uninterpretable.

\subsection{Discussion}

The results from Study 1 indicate that users on the Scratch online
community have a wide range of responses to remixing. People who responded were just
as likely to leave positive comments (21\%) as they were to leave a
direct or indirect complaint of plagiarism (22\%). These results
however, leave open the question of why such a wide range of comments
were left by originators. Several potential answers to these questions
are explored in Studies 2 and 3.

\subsection{Implications for design}

The Scratch online community was designed explicitly as a platform for
sharing and remixing media. Despite the fact the Scratch
infrastructure provided the technical facilities to remix content
easily, a set of explicit norms and licenses communicated to
users through links to a child-friendly version of the
pro-remixing license on every project, and continuous proselytizing of
remixing by the administrators of the site, many users reacted
negatively to remixes and expressed a sentiment that remixers had
plagiarized their work.

Creative Commons has described its work, both through the creation of
licenses permitting remixing and through the creation of technological
systems built around RDF (Resource Description Framework) metadata, as
means of reducing permission-asking \cite{lessig_keynote_2004}. Like
Scratch, many social media and remixing communities use CC's legal and
technological systems.  To the degree that our results generalize, our
findings suggest that the technical and normative permission to create
remixes may be insufficient to supporting positive reactions to remixing
in a social media remixing community.

\section{Study 2: When do originators accuse\\ remixers of plagiarism?}

\begin{table*}[ht]
\begin{center}

\begin{tabular}{rrrrrrrrrr}
  \hline
 & Mean & SD & 1 & 2 & 3 & 4 & 5 & 6 & 7 \\ 
  \hline
1. ACCUSE.PLAG & 0.13 & 0.34 &  &  &  &  &  &  &  \\ 
  2. SPRITES & 9.05 & 12.24 & 0.08 &  &  &  &  &  &  \\ 
  3. ORIG.REMIX & 0.05 & 0.21 & -0.01 & 0.00 &  &  &  &  &  \\ 
  4. HAS.REMIXED & 0.79 & 0.41 & -0.01 & -0.07 & 0.11 &  &  &  &  \\ 
  5. FEMALE & 0.25 & 0.43 & -0.06 & -0.11 & 0.04 & 0.11 &  &  &  \\ 
  6. WEEKS & 35.15 & 14.49 & 0.01 & -0.03 & 0.08 & 0.15 & 0.15 &  &  \\ 
  7. AGE & 17.08 & 10.15 & -0.07 & 0.08 & 0.01 & -0.08 & -0.20 & -0.16 &  \\ 
  8. REMIXER.AGE & 15.01 & 9.58 & -0.05 & 0.01 & 0.06 & -0.09 & -0.05 & -0.05 & 0.11 \\ 
   \hline
\end{tabular}

\caption{Means, standard deviations, and correlations between variables
used in the logistic regression analysis in Study 2. The sample includes
all remixed projects that had been clicked on and viewed by the
originators. ($n$=3742)}

\label{tab:cor}

\end{center}
\end{table*}

There are several explanations for the wide variety of reactions to
remixing shown in Study 1. In Study 2, we use the results of Study 1 to
construct a variable measuring whether originators have accused a
remixer of plagiarism.  We use this construct as the dependent
variable in a series of fitted logistic regression models to provide
initial tests of support for several explanations of why originators may
accuse remixers of their work of plagiarism using additional data on projects
and their creators.

One difference between Scratch and some other online peer-production
communities is that many Scratch projects are constructed at enormous
individual effort. On Scratch, users share full-fledged games or
animations with code, artwork and sound.
This is in contrast to many peer production communities, like
Wikipedia, where users usually contribute smaller portions of articles or
small fixes. One explanation for the high number of
complaints on Scratch may be that Scratch users develop a stronger sense
of ownership because of the large amount of individual time and effort
creators invest in their projects. This sense of ownership may set users
up to be more protective of their work and more likely to accuse
remixers of plagiarism.  This explanation leads us to our first
hypothesis (H2-1): {\em Originators of larger or more complicated
contributions will be more likely to accuse remixers of their projects
of plagiarism.}

In many peer production communities, like Wikipedia, the vast majority
of contributions are to existing products and work is primarily
cumulative in nature. In our sample of 11,861 remixes from the first year of Scratch's activity, the large majority of remixes (11,493 or 97\%) were based on
projects that were created {\em de novo} while the remaining were second
generation remixes. If users feel more protective of projects that are
entirely the product of their own work, an explanation for the high
number of complaints in Scratch is that because such a large number of
Scratch projects are created {\em de novo}, Scratch users are more
likely to feel plagiarized when their work is remixed. This leads to our
second hypothesis (H2-2): {\em Originators will be less likely to accuse
remixers of their projects of plagiarism when the remixed project is
itself a remix.}

A final explanation extends this reasoning from the project level to the
individual. Perhaps the process of creating remixes encourages
originators to be empathetic toward remixers and to integrate these
users into a ``remixing culture'' where copied or slightly modified
projects are not seen as plagiarism but rather as positive
contributions. In our sample, most authors of remixed projects were
active contributors who have, at some point in time, created their own
remixes.  Indeed, only 26\% of users in our sample (n=3,085) have never
shared a remix. It is possible that a large portion of charges of
plagiarism come from users who have not been integrated into Scratch's
``remix culture'' and are opposed to remixing in general.  This
explanation leads to to the formation of our third hypothesis
(H2-3): {\em Originators will be less likely to accuse remixers of their
projects of plagiarism if they have shared at least one remix
themselves.}

Of course, other factors are likely to have an important impact on
responses to remixing in Scratch that we feel it is necessary to control
for.  For example, charges of plagiarism may be due, in part, to the
large number of projects created by males in our sample (only 25\% of
originators were female) who may be more likely to accuse others of
plagiarism.  Additionally, the proportion of projects that are remixes
has increased over the life of our study. This suggests that attitudes
toward remixing may have changed over time with the potential for a
change in plagiarism accusation rates.  Finally, younger originators may
be more likely to accuse remixers of plagiarism either because they do
not understand that they are giving permission to others to remix by
sharing their work, because younger users are more likely to react
negatively in general, or for any number of other factors that correlate
with age.  As a result, we will control for gender, the time period when
projects were shared, and age, when testing the above hypotheses.

\subsection{Procedure}

To explore these issues, we used a sample consisting of the 3,742
remixed projects that had been clicked on and viewed by the originators,
as described in Study 1.  Viewing a project is both a very low bar for
involvement in the community and a prerequisite for any type of response
to remixing -- the subject of our study -- even if that response is a
decision to not act. 


The Scratch online community is run using a custom built web application
with data stored in a MySQL database; we collected data for each of our
predictors from this source. Our dependent variable is a dichotomous
variable (ACCUSE.PLAG), constructed using our results in Study 1, which
measures whether originators accused remixers of plagiarism in a
non-positive manner. It is a dummy variable that takes the value of 1
when a comment was coded either {\em plagiarism} or {\em hinted
plagiarism} unless the comment is also coded {\em positive}. In our
discussion, we explain that we explored several alternative
specifications of this outcome with very similar results. 

Project complexity can be measured either through the amount of
programming code or the total number of graphical characters (SPRITES)
controlled by these scripts. Because these measures were highly
correlated (r=0.80) we choose to use SPRITES alone as our measure of
complexity. To aid in interpretation in our models below, we report
sprites in standard deviation units in our fitted models. To test
hypothesis H2-2 regarding the effect of cumulative contribution on
responses, we constructed a dummy variable (ORIG.REMIX) indicating if
the antecedent project was, itself, a remix. Similarly, we constructed a
dummy variable (HAS.REMIXED) indicating if the originator has ever
uploaded a remix, to test H2-3.

For our controls, gender is a dummy variable (FEMALE) indicating whether
the original author is female and was measured through self-reported
data from users' registration with the Scratch website. We measure time
based on upload data in the web application database. Because we had
reason to believe that the effect of time on plagiarism accusations may
be non-linear, we included the quadratic form of a variable measuring
the the number of weeks since the the first project was uploaded to the
live Scratch website (WEEKS).  We were able to measure the age of users
(AGE) through a self-reported birth month and birth year fields in the
Scratch registration for both the remixer and originator. We marked
age data as missing for users with ages under 4 and over 90 (139
observations for remixers and 124 for originators). Ages were calculated
at the day the remixed project was uploaded. Both ages are skewed toward
younger users with median values of 13 and 12 respectively --- several
years below the mean. A correlation table with means and standard deviations 
of all of the variables
included in our models is shown in Table \ref{tab:cor}.

\subsection{Results}

\begin{table*}[!ht]
\begin{center}

\begin{tabular}{ l D{.}{.}{3}D{.}{.}{3}D{.}{.}{3}D{.}{.}{3}D{.}{.}{3}D{.}{.}{3} } 
\hline 
  & \multicolumn{ 1 }{ c }{ Model 0 } & \multicolumn{ 1 }{ c }{ Model 1 } & \multicolumn{ 1 }{ c }{ Model 2 } & \multicolumn{ 1 }{ c }{ Model 3 } & \multicolumn{ 1 }{ c }{ Model 4 } & \multicolumn{ 1 }{ c }{ Model 5 } \\ \hline
(Intercept)            & -1.904 ^{***} & -2.159 ^{***} & -2.301 ^{***} & -2.302 ^{***} & -2.287 ^{***} & -2.155 ^{***}\\ 
                       & (0.049)       & (0.312)       & (0.315)       & (0.315)       & (0.325)       & (0.335)      \\ 
FEMALE      &               & -0.534 ^{***} & -0.496 ^{***} & -0.495 ^{***} & -0.494 ^{***} & -0.489 ^{***}\\ 
                       &               & (0.127)       & (0.128)       & (0.128)       & (0.129)       & (0.131)      \\ 
WEEKS     &               & 0.064 ^{***}  & 0.064 ^{***}  & 0.064 ^{***}  & 0.064 ^{***}  & 0.065 ^{***} \\ 
                       &               & (0.019)       & (0.019)       & (0.019)       & (0.019)       & (0.020)      \\ 
WEEKS$^2$  &               & -0.001 ^{***} & -0.001 ^{***} & -0.001 ^{***} & -0.001 ^{***} & -0.001 ^{***}\\ 
                       &               & (0.000)       & (0.000)       & (0.000)       & (0.000)       & (0.000)      \\ 
AGE               &               & -0.029 ^{***} & -0.031 ^{***} & -0.031 ^{***} & -0.031 ^{***} & -0.028 ^{***}\\ 
                       &               & (0.006)       & (0.006)       & (0.006)       & (0.006)       & (0.006)      \\ 
SPRITES (std)        &               &               & 0.210 ^{***}  & 0.210 ^{***}  & 0.209 ^{***}  & 0.205 ^{***} \\ 
                       &               &               & (0.042)       & (0.042)       & (0.042)       & (0.044)      \\ 
ORIG.REMIX      &               &               &               & -0.059        &               &              \\ 
                       &               &               &               & (0.247)       &               &              \\ 
HAS.REMIXED &               &               &               &               & -0.022        &              \\ 
                       &               &               &               &               & (0.124)       &              \\ 
REMIXER.AGE            &               &               &               &               &               & -0.018 ^{**} \\ 
                       &               &               &               &               &               & (0.007)       \\
 $N$                    & 3742          & 3615          & 3615          & 3615          & 3615          & 3480         \\ 
AIC                    & 2888.162      & 2758.412      & 2736.849      & 2738.790      & 2738.816      & 2599.331     \\ 
BIC                    & 2913.072      & 2882.269      & 2885.477      & 2912.190      & 2912.216      & 2771.665     \\ 
$\log L$              & -1440.081     & -1359.206     & -1344.424     & -1341.395     & -1341.408     & -1271.665     \\ \hline
 \multicolumn{7}{l}{\footnotesize{Standard errors in parentheses}}\\
\multicolumn{7}{l}{\footnotesize{$^\dagger$ significant at $p<.10$; $^* p<.05$; $^{**} p<.01$; $^{***} p<.001$}} 
\end{tabular} 

\caption{Taxonomy of logistic regression models on ACCUSE.PLAG, a
dichotomous construct representing whether project creators accused the
remixer of their project of plagiarism in a non-positive manner.}
\label{tab:reg}

\end{center}

 \end{table*}

Results of our fitted regression models are shown in Table
\ref{tab:reg}.  Model 0 is our unconditional model and Model 1 is our
control model which adds variables controlling for the originator's
gender, the quadratic term measuring weeks between the original project
upload and Scratch's launch, and the age of the originator. The effect
of originator gender on the likelihood of plagiarism accusations is
highly statistically significant, large, and stable across subsequent
specifications. Indeed, our model estimates that, robust to the addition
of all of the other controls in our model, the odds of a
female accusing a remixer of plagiarism is less than 0.6 times the odds
of males doing so. Both parameters in the quadratic terms measuring the
number of weeks since the projects were uploaded are statistically
significant and robust across specifications. Finally, our measure of
originator age also has an effect on the outcome that is statistically
significant and robust across subsequent specifications. Controlling for
gender, younger students are indeed more likely to accuse a remixer of
plagiarism. We tested for a quadratic age term and an interaction
between age and gender and found no statistically significant effect of
either on the outcome.  Although we are skeptical that the effect of age
on plagiarism accusation rates is linear, we suspect that this result is
a factor of our data which is largely limited to younger users where the
relationship may indeed be estimated as such.


Model 2 adds our measure of complexity, a variable measuring the number
of sprites in a project in standard deviation units, which we estimate
is associated with a higher likelihood of plagiarism accusations.  With
our controls in the model, we estimate in Model 2 that the odds that an
originator of a project will accuse a remixer of their project of
plagiarism are 1.23 times higher than the odds that the originator of a
project with one standard deviation (12.2) fewer sprites will do so.
Consequently, we find support for hypothesis H2-1 in that, even with the
addition of controls for age, gender, and when the project was posted,
originators are more likely to accuse remixers of plagiarism when the
remixed project is more complex.

Model 3 adds the dummy variable indicating whether the original project
in question is a remix itself. In our model, we do not find a
statistically significant effect of this dummy variable on our outcome.
In other words, we cannot reject the null hypothesis that originators
are as likely to accuse remixers of their projects of plagiarism when
the remixed project was itself a remix as when it was an original
production. As a result, we do not find support for hypothesis H2-2
that, controlling for gender, age, and project complexity, originators
will be less likely to accuse remixers of their projects of plagiarism
when the remixed project is itself a remix.

Model 4 instead adds to Model 2 the dummy variable reflecting whether
originators have ever uploaded a remix themselves. Once again, we do not
find a statistically significant effect of this predictor on the
outcome. As a result, we also fail to find support for our final
hypothesis H2-3 that, controlling for gender, age, and project
complexity, originators who have uploaded remixes themselves are less
likely to accuse remixers of their work of plagiarism.

As a robustness check, we re-estimated our models on a data set that
excluded projects shared before May 15, 2007, the first day that
widespread press reports of the Scratch community were disseminated. In
the period before, users were a smaller subset
who may have been more likely to know each other in person. Our results
were not substantively affected. We also estimated models on a data set
that did not exclude implausibly high and low ages, and found that our
results were similar once again.

We also estimated our models using slightly different alternate
specifications of our dependent variable. Because many negative
reactions by originators are due to plagiarism but do not explicitly
call it out, we used a specification of our dependent variable that was
also true for negative reactions that did not specify plagiarism with
very similar results. We also reformulated our dependent variable so it
only included explicit charges of plagiarism that were not paired with
positive messages (i.e., {\em hinting plagiarism} charges were not
included). Our results were, once again, substantively unchanged.



\subsection{Discussion}

We found support for the theory that creators are more likely to accuse
remixers of plagiarism if the remixed project is more complex.  To the
degree that our results generalize to other online communities, charges
of plagiarism may be of reduced concern in communities where individual
contributions tend to be small.

Surprisingly, our models suggest no effect of whether the project was itself a
remix on the rate of plagiarism accusations. This might indicate that Scratch users accusing remixers of
plagiarism have a strong conception of ``good'' (e.g., original and
transformative) remixes and ``bad'' (e.g., plagiarizing) remixes which
are simple copies. In line with this explanation, we did not find
support for hypothesis H2-3, originators who have uploaded remixes were
neither more nor less likely accuse remixers of plagiarism than users
who had never uploaded a remix.

Although included as a control, the effect of age suggests intriguing
future research. Future work could be designed to address why
younger children may be more likely to complain about plagiarism.
For example, one possible explanation is that young remixers do not understand
licensing.  On the other hand, previous qualitative work suggests that,
although significant, other factors may put important limits on the
understanding by or desire of users to pay attention to licenses. For
example, Diakopoulos et al.  showed that adult users on an online video
sharing site asked for permission before reusing media, despite
licensing considerations which made it clear that such use was legally
permissible. Of course, other factors associated with age may also play
an important role in the relationship we observe.

Finally, while our framing and the
variables in our model attempt to capture aspects of originators and
their projects which may affect the probability of originators accusing
remixers of plagiarism, aspects of remixers and their remixes almost
certainly play an important role in setting up projects for negative
feedback by the author of an antecedent.

The high correlation between qualities of remixes and their antecedents
makes exploring this comparison difficult in our data set. As one simple
effort to probe this explanation, we offer Model 5 (shown in Table
\ref{tab:reg}) as an example which adds a variable to Model 2 that
measures the age of the remixer at the time of the remixes' upload.
Controlling for originators' gender, the date, originators' age, and the
complexity of the remixed project, we estimate that remixes by younger
users are more likely to result in accusations of plagiarism. Of
course, as discussed above, the effect of age on our outcome is
difficult to interpret reliably alone.  However, even as a tentative
result, this model provides support for the argument that accusations of
plagiarism are influenced by what each remix consists of, and by who the
remixer is, as well as by aspects of the person leaving the feedback. We
make a further attempt to unpack these results in Study 3.

\subsection{Implications for design}

In Study 1, we showed how a technical capacity to remix and normative
statements in support of remixing do not guarantee either positive
reactions or an elimination of charges of plagiarism. In Study 2, we
unpack our initial results, and evaluate several explanations of the
difficulties that designers may encounter when attempting to address
these problems.

Our findings support the theory that the importance of systems to
address charges of plagiarism may be higher in communities where
contributions are smaller. Even within Scratch, where every contribution
is in the form of a stand-alone project, differences in project
complexity are associated with large differences in the probability that
an author will accuse a remixer of plagiarism. Although we cannot speak
to causal effects, to the degree that our results generalize to other
communities, our findings imply that encouraging cumulative contribution
may not result in a lower rate of plagiarism accusations.  Although
designers may be encouraged to involve more users in remixing as a way
of increasing positive attitudes toward remixing, the relationship might
be more complex or less tightly associated than some designers might
assume.  Scratch's example suggests that increased participation in
remixing alone may not correspond to a decreased likelihood of plagiarism
accusations.

\section{Study 3: Are plagiarism complaints more common when remixes are
more similar?}

While the technical and legal ability to remix is constant across
the Scratch online community, the nature and content of remixes vary
extensively. Some remixes are near or even prefect copies of the project
they are based on while others are extensive remixes that bear little
similarity.

In Study 2, we explored several explanations for the high number of
complaints by focusing on qualities of originators and of the remixed
project. Of course, as we alluded to in our discussion of Study 2,
originators' reactions to remixes are also likely to be influenced by
the nature of the remix and the remixer. Perhaps the most obvious
remixer-side explanation for the wide range of responses to plagiarism
in Study 1 is that the extent to which remixers rely on the original
project varies.  That is, users may not mind remixing when the remix is
merely inspired by or loosely based on their work but object when
the remixed project is nearly identical to their own.  Study 3 makes a
first attempt to investigate this hypothesis (H3): {\em Originators are
more likely to accuse remixers of plagiarism when the remixed project
and its antecedent are more similar to each other.}

Because qualities of remixes are highly correlated with qualities of
remixed projects, adding remix-level variables to our logistic
regression model in Study 2 was untenable with our data set and
methods.  Similarly, automatic methods of measuring differences between
remixes and their antecedents available to use were found to be unreliable.
Hand-coding is possible but requires viewing and interacting with each
pair of projects and is extremely time intensive. As a result, Study 3
represents a first attempt to explore project similarity by offering a
bivariate comparison between originator reactions and project similarity
using a reduced, non-representative, sample.

\subsection{Procedure}

A random selection of 40 originator-remixer project pairs from each of
the 6 categories of comments (e.g., plagiarism, hinting plagiarism, no
comment, etc.; total n=240) were put in a random order and were given to
a new pair of coders who were unaware of how these projects were
selected, that these projects represented six categories of projects, or
that their selection had anything to do with the comments left on these
projects. These coders were asked to watch and/or play each of the
projects in each pair and to make a judgment of similarity on a 5-point
scale (from 1={\em can't tell they are related} to 5={\em can't tell
they are different}). Their responses were highly correlated,
(r=.79, p$<$.001; Cronbach's $\alpha$=.88), they rated them within
one point of each other in 95\% of cases, and these ratings were
averaged for a final similarity score for each project pair.

\subsection{Results}

We conducted a one-way ANOVA on similarity ratings as a function of the
type of comment left. Similarity influenced the type of comment left
(F$_{(5,234)}$=4.78, p$<$.001).  Since we were specifically interested
in assessing whether more similar projects were more likely to lead to
plagiarism concerns, we conducted planned contrasts,  comparing the similarity scores of the {\em
plagiarism} ($\mu$=4.40, $\sigma$=.65) group with scores in the other
groups -- {\em doesn't fit} ($\mu$=3.53, $\sigma$=.85), {\em negative}
($\mu$=3.93, $\sigma$=1.02), {\em positive} ($\mu$=3.75, $\sigma$=.85),
{\em hinting plagiarism} ($\mu$=3.46, $\sigma$=1.30) and {\em no
comment} ($\mu$=3.65, $\sigma$=1.14) groups. This analysis revealed that
accusations of {\em plagiarism} were associated with more similar
remixes than the {\em hinting plagiarism} projects (t$_{234}$=4.22,
p$<$.001, d=0.91), the {\em doesn't fit}  projects (t$_{234}$=3.94, p$<$.001, d=1.15),
{\em no comment} projects (t$_{234}$=3.38, p=.001, d=0.81), {\em positive}
projects (t$_{234}$=2.93 , p=.004, , d=0.85), and  {\em negative} projects
(t$_{234}$=2.14, p=.033, d=0.55).

\subsection{Discussion}

This study indicated that plagiarism accusations were influenced by the
similarity between the original work and the remix and these findings
give tentative support for H3. When remixes were highly similar to the
original projects, they were much more likely to elicit an accusation of
plagiarism.

\subsection{Implication for design}

Designers of social media remixing systems may be able to decrease
charges of plagiarism by remixed users by promoting differentiation
between remixed projects and their antecedents. In particular, users
might react more positively if a system either created technical
affordances to create dissimilar remixes or to highlight differences
between apparently similar projects.

For example, in Scratch, remixers begin with an unmodified version of
the full source of the project to be remixed. An example of technical
affordances to facilitate differentiated project might be a remixing
interface that begins with a blank project and treats remixed projects as
sources for code and media. However, such design affordances may present
negative consequences in other areas of the site by increasing the cost
to users of making simple improvements or engaging in more direct forms
of collaboration. Another suggestion for Scratch may
include a ``changelog'' facility that allows users to explain
substantive differences between a remix and an apparently similar
antecedent project. For example, a user who fixes a bug or changes a set
of sprites could explain initially unnoticeable changes. By emphasizing
differences, both apparent similarity and charges of plagiarism might be
decreased.

\section{Conclusions}

These studies explore attitudes toward remixing that we believe are important.  In Study 1, we show that
users react to remixing in a wide variety of ways. Although every
project on Scratch is shared under a license that permits remixing, as
many authors of original projects accuse remixers of plagiarism as react
positively. Study 2 tested three hypotheses about aspects of antecedent
projects and their creators that might be related to reactions to
remixing. Although our analysis cannot offer causal explanations, our
findings support the theory that the authors of more complex projects
tend to accuse others of plagiarism at a higher rate.  On the other
hand, we do not find support for the hypotheses that authors of original
projects that are themselves remixed, or authors who have never
published remixes accuse remixers of plagiarism at a higher rate. In
Study 3, we present tentative findings that support the explanation that
users are more likely to make accusations of plagiarism when projects
are more similar.

In this analysis, we make a number of critical assumptions. In general,
our framing tends to treat charges of plagiarism as negative and to be
avoided. This interpretation is roughly supported in our data set: 32\%
of comments coded as negative were also coded as explicitly calling out
plagiarism while only 2\% of positive comments did so. Of course, this
does not mean that charges of ``copy-cat'' are necessarily associated
with either bad feelings by users or, more importantly, behaviors that
social media designers find problematic. Although our understanding of
the coded comments and our experience with the community give us
confidence in our framing, further work should unpack these assumptions.

Indeed, promising future work might use attitudes toward and responses
to remixing as an independent variable. For example, designers of remixing
communities may want to look at the effect that reactions to remixing
have on the rate or nature of contributions. It seems unlikely that a
community hostile toward remixing or actively involved in calling each
other ``copy-cats'' would be solid foundation on which to build such a
culture. Future work will be able to build on the findings in this paper
to establish how these attitudes help frame a social environment.
Similarly, such work should look at the effect of positive reactions.
Although Study 2 focused on charges of plagiarism, positive responses
seem as likely to have an effect on remixing rates as negative reactions
and accusations of plagiarism. Future work should build on the work in
this paper to do so.


\section{Acknowledgements}
The authors wish to thank Alex Shaw for useful feedback on a previous draft of this manuscript. Scratch is a project of the Lifelong Kindergarten Group at the MIT Media Lab with financial support from the National Science Foundation (Grant No. ITR-0325828), Microsoft Corp., Intel Foundation and the MIT Media Lab research consortia.

\bibliography{icwsm_paper.bib}
\bibliographystyle{aaai}
\end{document}